\documentclass[journal]{IEEEtran}
\usepackage{float}
\usepackage{subfigure}
\usepackage[justification=centering]{caption}

\usepackage{ifpdf}

\usepackage{cite}

\ifCLASSINFOpdf
\usepackage[pdftex]{graphicx}
\else
\usepackage[dvips]{graphicx}
\fi
\usepackage{subfigure}

\usepackage{amsmath}
\usepackage{array}
\usepackage{algorithmic}
\usepackage{fixltx2e}

\usepackage{stfloats}

\usepackage{url}
\usepackage{booktabs}

\usepackage{multirow}
\usepackage{chngpage}
\usepackage{hyperref}
\usepackage{amssymb}
\usepackage[justification=justified]{caption}
\usepackage{float}
\usepackage{stfloats}
\usepackage{xcolor}
\usepackage{breakurl}
\hypersetup{
	colorlinks   = true,
	citecolor    = blue,
	linkcolor = blue,
	anchorcolor = blue,
	filecolor = blue,
	linkbordercolor = blue,
	urlcolor = blue
}
\usepackage{hyperref}

\usepackage[misc]{ifsym}

\begin{document}
	
	\title{Unsupervised Landmark Detection Based Spatiotemporal Motion Estimation for 4D Dynamic Medical Images}
	\author{Yuyu Guo\textsuperscript{1}, \and
		Lei Bi\textsuperscript{2(\Letter)}, \and
		Dongming Wei\textsuperscript{1}, \and
		Liyun Chen\textsuperscript{1}, \and
		Zhengbin Zhu\textsuperscript{3}, \and
		Dagan Feng,\textsuperscript{2}~\IEEEmembership{Fellow,~IEEE},
		Ruiyan Zhang\textsuperscript{3}, \and
		Qian Wang\textsuperscript{1(\Letter)} and
		Jinman Kim\textsuperscript{2(\Letter)}
		
		\thanks{Y. Guo, D. Wei, L. Chen and Q. Wang are with the School of Biomedical Engineering, Shanghai Jiao Tong University, Shanghai 200030, China, e-mail: wang.qian@sjtu.edu.cn.}
		
		\thanks{L. Bi, D. Feng and J. Kim are with the School of Computer Science, University of Sydney, NSW 2006, Australia, e-mail: lei.bi@sydney.edu.au, jinman.kim@sydney.edu.au.}
		
		\thanks{Z. Zhu and R. Zhang are with Ruijin Hospital, Shanghai Jiao Tong University School of Medicine, Shanghai 200025, China.}
		
		
		
		
		
		\thanks{This work was supported in part by Australian Research Council (ARC) grants (LP140100686 and IC170100022), the University of Sydney – Shanghai Jiao Tong University Joint Research Alliance (USYD-SJTU JRA) grants.}
	}
	
	%
	%

	\markboth{Journal of \LaTeX\ Class Files,~Vol.~14, No.~8, August~2015}%
	{Shell \MakeLowercase{\textit{et al.}}: Bare Demo of IEEEtran.cls for IEEE Journals}
	%



	\maketitle
	\thispagestyle{empty}
	\begin{abstract}
		Motion estimation is a fundamental step in dynamic medical image processing for the assessment of target organ anatomy and function. However, existing image-based motion estimation methods, which optimize the motion field by evaluating the local image similarity, are prone to produce implausible estimation, especially in the presence of large motion. In addition, the correct anatomical topology is difficult to be preserved as the image global context is not well incorporated into motion estimation. In this study, we provide a novel motion estimation framework of Dense-Sparse-Dense (DSD), which comprises two stages. In the first stage, we process the raw \textit{dense} image to extract \textit{sparse} landmarks to represent the target organ’s anatomical topology, and discard the redundant information that is unnecessary for motion estimation. For this purpose, we introduce an unsupervised 3D landmark detection network to extract spatially \textit{sparse} but representative landmarks for the target organ’s motion estimation. In the second stage, we derive the sparse motion displacement from the extracted \textit{sparse} landmarks of two images of different time-points. Then, we present a motion reconstruction network to construct the motion field by projecting the sparse landmarks' displacement back into the dense image domain. Furthermore, we employ the estimated motion field from our two-stage DSD framework as initialization and boost the motion estimation quality in light-weight yet effective iterative optimization. We evaluate our method on two dynamic medical imaging tasks to model cardiac motion and lung respiratory motion, respectively. Our method has produced superior motion estimation accuracy compared to existing comparative methods. Besides, the extensive experimental results demonstrate that our solution can extract well representative anatomical landmarks without any requirement of manual annotation. Our code is publicly available  \textcolor{blue}{\hyperref{https://github.com/guoyu-niubility/DSD-3D-Unsupervised-Landmark-Detection-Based-Motion-Estimation}{code}{DSD}{Online\footnote{\hyperref{https://github.com/yyguo-sjtu/DSD-3D-Unsupervised-Landmark-Detection-Based-Motion-Estimation}{code}{DSD}{https://github.com/yyguo-sjtu/DSD-3D-Unsupervised-Landmark-Detection-Based-Motion-Estimation}}.}} 
	\end{abstract}
	
	\section{Introduction}

	
	Four dimensional (4D) dynamic medical imaging is a vital technique for organ motion monitoring, as it provides comprehensive spatial temporal information via multiple 3D volumes sampled over a period of time \cite{kwong2015f}. These 4D images can be used to estimate motion of the target organ which allows for the assessment of the functional and structural properties of target organs that aid in clinical outcomes. Examples of motion estimation are abundant, including functional heart analysis with 4D-MRI \cite{bornstedt2001multi}, respiratory organ motion modelling with 4D-CT \cite{pan20044d}, image-guided intervention with 4D-CT \cite{cleary2010image}, diagnosis of disease and treatment planning \cite{fernandes2020reliable}.
	
	\begin{figure}[h]
		\begin{center}
			\includegraphics[width=0.8\linewidth]{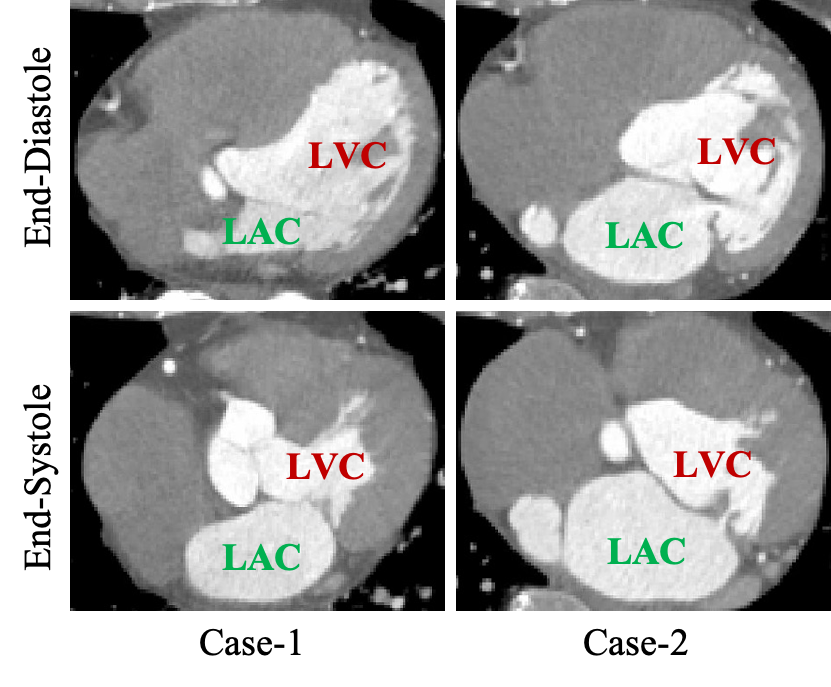}
		\end{center}
		\caption{Examples of 4D cardiac imaging of two patient cases, respectively of the two columns. The temporal variation between the image sequence can be large, i.e., between end-systole (ES) and end-diastole (ED) shown in two rows. All images are shown in trans-axial views and cropped to focus on the heart. LVC indicates the left ventricle cavity, while LAC indicates the left atrium cavity.}
		\label{fig:1}
	\end{figure}
	
	
	Technically, motion estimation is to establish corresponding deformation field between two time-point images which can be attained by image registration methods \cite{buerger2011hierarchical,yu2020foal}. \textcolor{black}{Image registration is a broad topic and widely applied in various fields, e.g., image reconstruction \cite{deng2020line,xuan2021multi}, image super-resolution \cite{zhang2018longitudinally}, and atlas-based segmentation \cite{wu2013estimating,zhuang2018multivariate}, etc. Particularly, in our prior work \cite{guo2020spatiotemporal}, we presented a supervised interpolation network leveraging image registration to improve the sequential sampling rate for individual subjects.} Conventional image registration approaches derive the motion field between two input images (usually referred to as the fixed image and moving image) by solving an optimization problem, which aims to maximize the image appearance similarity under certain smoothness regularization \cite{vercauteren2009diffeomorphic,avants2011reproducible}. However, these methods are typically time-consuming (or computationally expansive) via the iterative optimization process. Besides, when coping with large deformation,  such as in cardiac motion between the two ends of the motion cycle (end-systole (ES) and end-diastole (ED), as shown in Fig. \ref{fig:1}), non-physiological misestimation may happen as the price of excessive pursuing of appearance similarity.

	Recently, deep-learning-based registration methods draw much attention due to their superior efficiency \cite{haskins2020deep}. They have proved that the conventional optimization can be replaced by a single deep model. These methods can be performed in an unsupervised learning manner \cite{de2019deep,ghosal2017deep}. They typically input a pair of images to the convolutional neural network (CNN), and then encode the appearance features hierarchically to decode the deformation field between the input images. Nevertheless, the performance of CNN in registration is limited by the receptive field of the convolution kernels, which may lead to underestimation in the presence of large deformation.

	Apart from image-based registration approaches, anatomical prior, such as landmarks or segmentation masks, is another crucial branch to aid motion estimation \cite{sotiras2013deformable,liu2019weakly}. These anatomical prior approaches firstly extract the sparse topological representation (\textit{i.e.}, landmarks) of the target organ, other than using original image in the \textit{dense} \textcolor{black}{representation form }\cite{xu2017local}. Then, the motion field is predicted via the \textcolor{black}{corresponding} landmarks between two different volumetric images. Compared with estimating directly from the images, these anatomy-based methods are usually accurate and robust \cite{viergever2016survey} when coping with large deformation. However, the drawback is that user interaction is often required to annotate the representative landmarks or segmentation masks, which is prone to manual errors and admittedly labor-intensive. For solving this issue, some studies present automatic identification of landmarks (keypoints) \cite{yu2014key,huang2014hand}. \textcolor{black}{These methods extract the keypoints by calculating the image local features and establish the mapping relationship between the keypoints to obtain the motion information of the target objects.} Even so, these methods still require the point matching step to compute the sparse displacement of the landmarks, and then reconstruct the dense motion field in the image domain \cite{li2014nonrigid,yu2019joint}. This is likely to increase the risk of errors occurring in landmark detection and matching, which may reduce the accuracy of the final motion estimation. Moreover, these methods lack demonstration of their efficiency in volumetric medical images.

	To estimate the accurate motion in dynamic 4D medical images, especially for the images with large deformation, we present a novel learning-based motion estimation framework of Dense-Sparse-Dense (DSD). In summary, there are two main contributions in our proposed DSD framework:
	\begin{enumerate}
		\item We introduce a motion estimation framework based on the guidance of the sparse topology change. In our framework, the motion field is estimated from the dynamic changes of the target organ in its sparse landmark representation. This is to reduce the error motion estimation in the existence of large deformation by leveraging the topological guidance.
		\item We propose a 3D unsupervised landmark detection network for volumetric medical image. It can extract effectively sparse landmarks that represent the anatomical topology of the target organ without any manual annotation. Specifically, we design multiple loss functions to guide the anatomical landmark detection network and utilize optimal transport theory to enforce prior topology constraint to enhance the richness of landmarks' anatomical description. 
	\end{enumerate}
	Benefiting from the guidance of the target organ's anatomy, the estimated motion field can effectively improve its plausibility and better cope with large deformation. We evaluate our proposed framework on two clinical motion analysis tasks - cardiac motion and lung respiratory motion - and achieve outstanding performance. 
	

	\section{Related Works}
	
	We partition the related works into three topics: (1) image-based registration, (2) landmark-based registration, and (3) landmark detection. 
	
	\subsection{Image-based Registration}
	\textcolor{black}{Image registration is a basic tool in motion estimation, which can be mainly divided into two categories, \textit{i.e.}, non-learning-based methods and learning-based methods. Classic non-learning-based approaches mainly include diffeomorphic Demons \cite{vercauteren2009diffeomorphic}, Free-Form Deformation (FFD) \cite{modat2010fast}, \textit{etc.} These methods derive the deformation field by iteratively maximizing the image similarity under certain smoothness regularization. Both methods are widely used in medical image registration, and are often regarded as the comparing baselines in related researches. However, the hyper-parameters of them are not easy to tune, while tremendous optimization costs high in computation time.}

	Inspired by the great success of deep learning, several learning-based studies have been proposed in recent years. Many of them leverage the supervised learning strategy, which requires the ground-truth (deformation field) derived from conventional registration tools to train the model. For instance, Yang et al. \cite{yang2017quicksilver} utilized a U-Net architecture \cite{ronneberger2015u} to predict the deformation field for brain MR images. Similarly, Rohe et al. \cite{rohe2017svf} also leveraged U-Net to generate the motion field for cardiac volumes. \textcolor{black}{Krebs et al. \cite{krebs2017robust} present a supervised agent learing based non-rigid registration method to learn the deformation field in low-dimension representation.} However, the supervision can be cumbersome to acquire, which restricts the deformation types to be learned.
	\begin{figure*}[h]
		\begin{center}
			\includegraphics[width=1\linewidth]{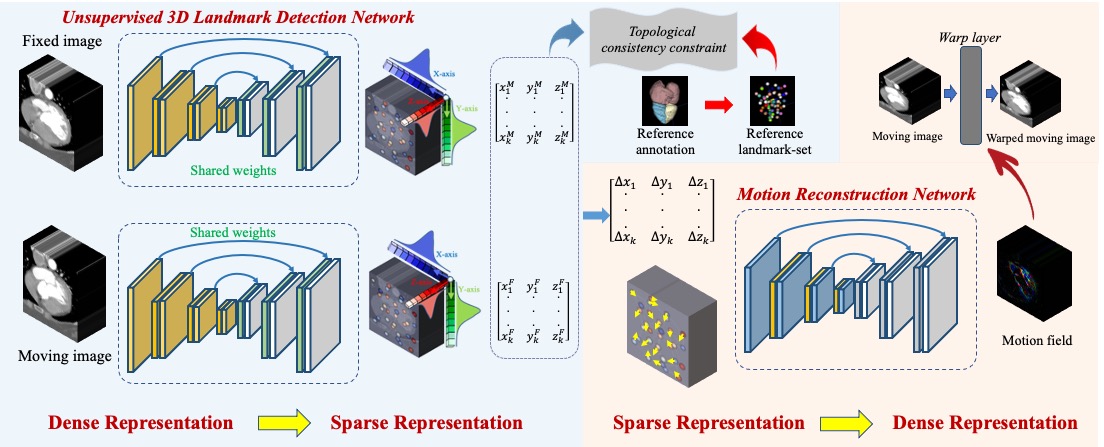}
		\end{center}
		\caption{An overview of our DSD framework which contains an unsupervised landmark detection network and a dense motion reconstruction network. The first process (with blue background) involves the fixed image and the moving image which are separately input to the landmark detection network to obtain their respective landmarks. In the second process (with orange background), we leverage a dense motion reconstruction network to estimate the dense motion field from the sparse motion vectors of the corresponding landmarks of the two input images.}
		\label{fig:2}
	\end{figure*}
	
	Recently, unsupervised learning solutions for registration have been proposed to eliminate high reliance on the ground-truth. Li et al. \cite{li2017non} and Vos et al. \cite{de2017end} introduced fully convolutional networks to estimate the deformation field for brain volumetric images and 4D cardiac images, respectively. Guha et al. \cite{balakrishnan2019voxelmorph} developed the popular VoxelMorph method for brain image registration. Zhang et al. \cite{zhang2018inverse} proposed an inverse-consistency loss based on a similar network of VoxelMorph. These unsupervised registration methods utilize a spatial transform layer with an image similarity loss to train the deep model. While the existing methods usually derive the dense \textcolor{black}{deformation} field over the entire image domain, the performance is prone to degrade for the situations such as large deformation.

	\subsection{Landmark-based Registration}
	Another category of registration is based on sparse landmarks \cite{jia2015novel,worz2014spline}. Landmarks are salient points of geometrical and anatomical uniqueness. Basically, the landmark-based registration methods are mainly applicable to estimate rigid or affine transformation. With the raising number of landmarks, they can also be used to describe more complex (\textit{e.g.}, non-rigid) deformation \cite{stewart2003dual}. Note, the identification of landmarks is particularly important for these methods. 
	
	The landmarks can be manually or automatically detected. Most existing landmark-based registration methods rely on manually labeled landmarks. One may directly derive the sparse displacement from the manually labelled landmarks, and further obtain the dense motion field with the help of interpolation such as thin plate spline (TPS) \cite{wood2003thin}. However, the manual labeling of landmarks is inevitably time-consuming and prone to intra- and inter-observer uncertainty, which limits its real application. \textcolor{black}{For automatic keypoint detection methods, they generally capture the keypoints through the image local features \cite{bi2021genetic}. For instance,  Yang et al. \cite{yang2018multi} presented a CNN feature-based motion tracking method for temporal images. Hosseini et al. \cite{hosseini2021non} utilize feature points to estimation the cardiac motion problem.} \textcolor{black}{Li et al. \cite{li2017multiview} introduced a Multiview-based Parameter Free framework, where pixel-wise motion estimation has been used to derive feature points for identifying the number groups of people in a crowd video. In another study, Liu et al. \cite{liu2017better} presented a point trajectories generation method by exploring the motion boundaries for video processing.} However, the automatic landmark detection usually comes with complex point matching \cite{8000598}, such as iterative closest point \cite{chetverikov2002trimmed}, to track the movement of corresponding landmarks and then to complete the following image registration. Although it does not require manual annotation, the errors caused by automatic landmark detection and subsequent point matching limit the accuracy and robustness in motion estimation, especially for medical images with large motion.

	\subsection{Landmark Detection}
	
	Landmark detection is widely used in computer vision, \textit{e.g.}, for human face/pose recognition \cite{zhu2016unconstrained,fang2017rmpe,huang2018clothing}. Initially, regression-based CNN methods utilize a fully convolutional layer to estimate the distinct positions. For instance, Yu et al. \cite{yu2016deep} presented a deformable landmark regression network that utilized a TPS layer to achieve the cross-talk between point-based deformation and images. Lv et al. \cite{lv2017deep} introduced a regression CNN with global stage and local stage to improve the performance on facial landmark detection. 
	
	Different from the regression approaches, localizing the landmark by predicting its heatmap is an alternative way. These methods usually use a kernel function to convert the coordinate information into \textcolor{black}{the form of heatmap}. Nie et al. \cite{nie2018hierarchical} proposed a hierarchical contextual refinement network for landmarks to efficiently predict the human pose. Siarohin et al. \cite{siarohin2019first} recently presented a heatmap-based landmark detection method to transfer the animation from video to still images. However, compared to exploring discriminate landmarks in 2D space, it is admittedly much more challenging for 3D landmark detection concerning the exponential growth in data complexity. Besides, compared with 2D facial images, 3D/4D medical images usually lack enough visual saliency, as the anatomical landmarks are severely interfered by noisy background.

	\section{Method}

	We aim to accurately estimate the motion field in 4D dynamic medical images with large motion, based on the guidance of the target organ's topological representation. As presented in Fig. \ref{fig:2}, an unsupervised landmark detection network is firstly utilized to recognize the discriminative and consistent landmarks from the temporal volumetric images. Then, we construct the motion field by projecting from the motion displacement of the \textit{sparse} landmarks back into the \textit{dense} motion field.

	Let $Q$ be the set of \textit{N} subjects. Each subject includes \textit{T} time-point (phase) images, denoted as \textit{$\{I_{t}\mid_{t=1...T}\}$}. The spatial domain can be denoted as $\varOmega$ for all subjects, where the image size is $(X, Y, Z)$. \textit{$\mathcal{H}_{t}$} denotes the predicted landmark heatmaps corresponding to the image \textit{$I_{t}$}. \textcolor{black}{We define the grid coordinate of the volumetric image as $\mathcal{G}_{(x,y,z)}$. The size of $\mathcal{G}_{(x,y,z)}$ is $(3, X, Y, Z)$, where $(X, Y, Z)$ represent the image size and the coordinate for each voxel is represented in three numbers.} The distinct locations of the individual landmarks for the image $I_{t}$ are denoted by $L_{t}=\{L_{tk}\mid_{k=1...K}\}$ where \textit{K} indicates the number of detected landmarks in each volumetric image. The motion field between the input images $I_{t_{m}}$ and $I_{t_{n}}$ is denoted as $\varphi_{t_{m} \rightarrow t_{n}}$, where $t_{m}$ and $t_{n}$ denote two time points. Note that the image sequences are all in the same temporal sampling.
	
	\subsection{From Dense Volumetric Image to Sparse Landmarks - Unsupervised Landmark Detection}\label{pretrain}
	
\begin{figure}[!htbp]
		\centering
		\includegraphics[width=0.5 \textwidth]{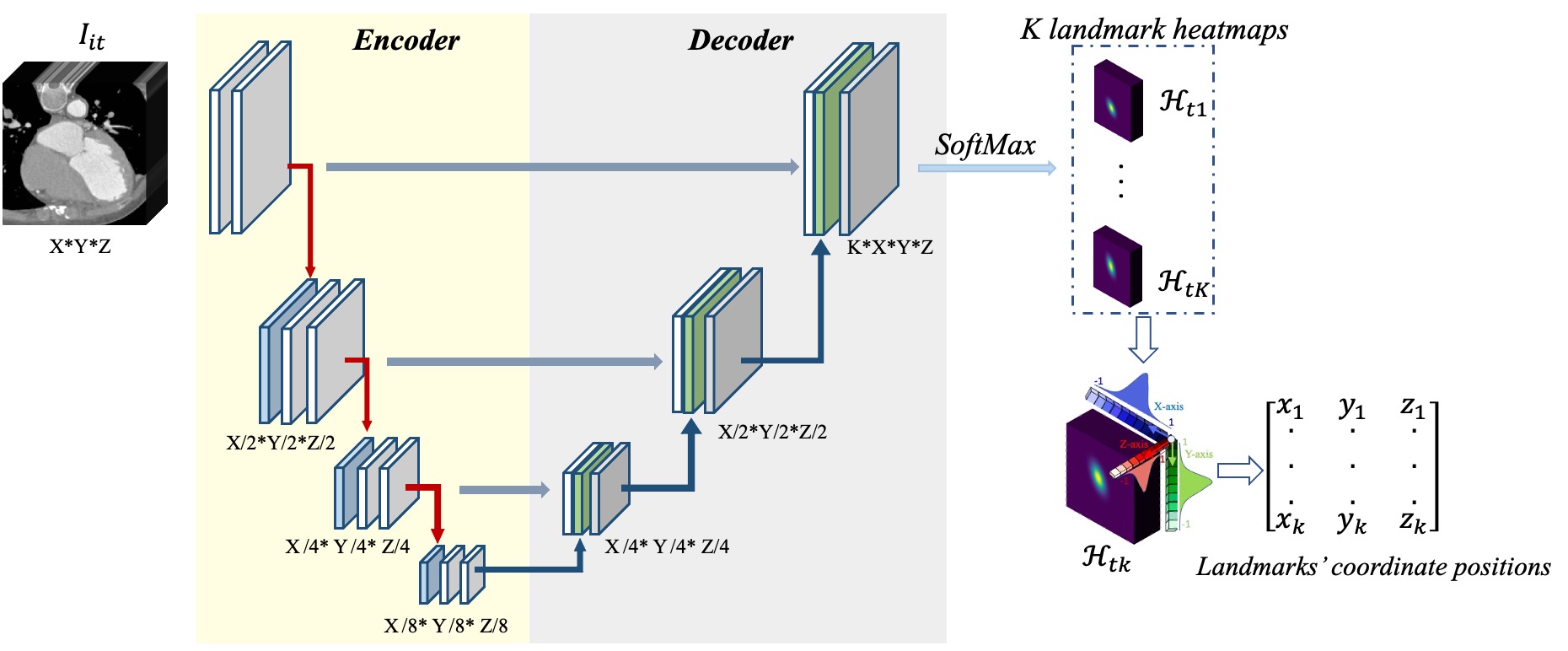} 
		\caption{The architecture of our 3D landmark detection network. The coordinate of the detected landmarks can be derived from the output heatmap.}
		\label{fig:4}
	\end{figure}

	The landmark detection serves as a core step in our \textit{DSD} motion estimation framework. To perform the landmark detection from \textit{dense} volumetric image, we propose an unsupervised learning-based landmark detection network. A standard multi-scale encoder-decoder architecture is used as the backbone, as shown in Fig.~\ref{fig:4}. To obtain representative landmarks \cite{vincent2010stacked}, we firstly introduce a self-supervised pre-training strategy to help the encoder of the landmark detection network to focus on the potential motion-affected regions. Then, for obtaining accurate and consistent landmarks, multiple losses are introduced to help train the network to perform landmark localization. Note that none of the annotation information is used to supervise any training step.
	
	\subsubsection{Self-Supervised Pre-Training for Motion Attention}
	The detected landmarks are required to represent the \textcolor{black}{temporal} changes of the target organ to further guide the motion estimation. To this end, we leverage a self-supervised pre-training strategy, to help the encoder of landmark detection network focus its attention toward the motion-affected regions. Considering that many organ/tissue movements of interest are in cycles (such as heart beating, breathing, \textit{etc.}), half-cycle movement is presumably monotonous, implying one-way morphological change from large volume to small (or inversely). Such characteristic can be utilized as the self-supervision to train the encoder network to recover the sequence order from randomly permuted one. In this way, the latent feature maps produced by the encoder can focus on the potential motion-affected regions in the temporal sequence.
	
	\begin{figure}[t]
		\centering
		\includegraphics[width=0.4 \textwidth]{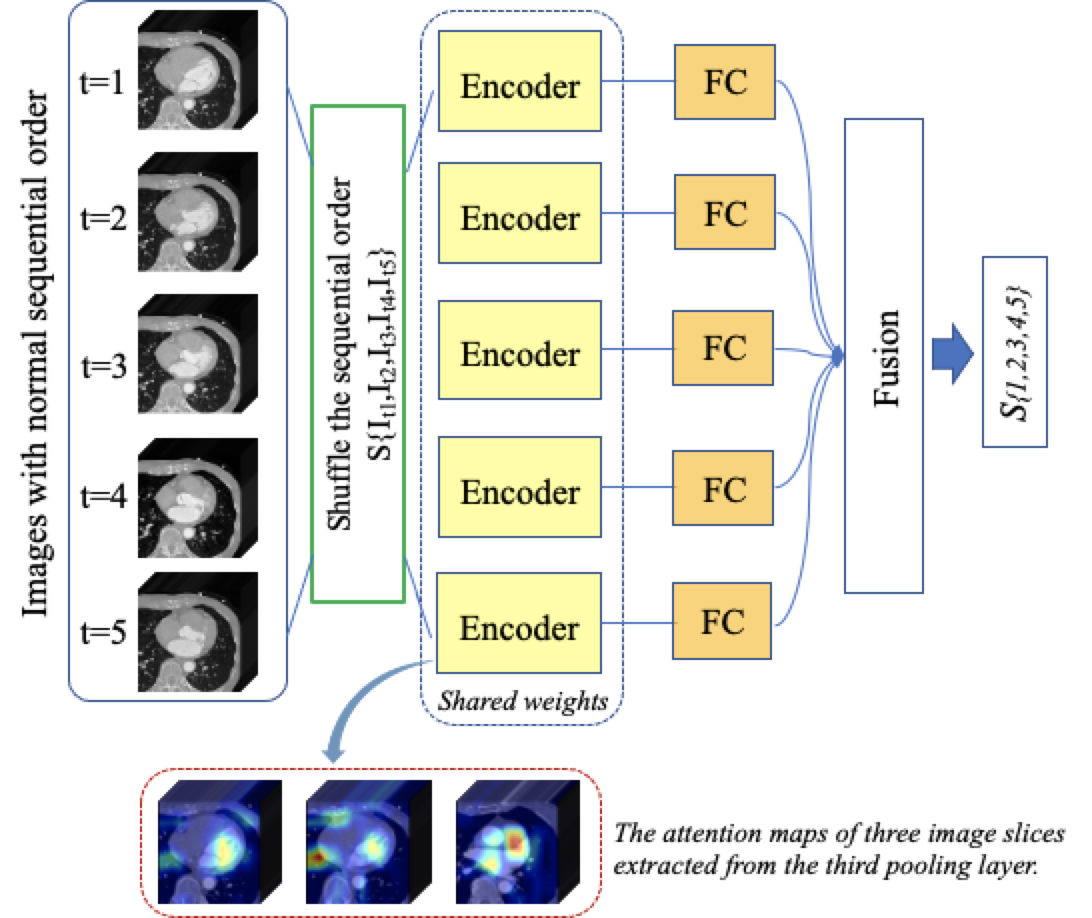} 
		\caption{A self-supervised learning strategy to obtain a well pre-trained encoder for landmark detection.}
		\label{fig:3}
	\end{figure}
	
	As depicted by Fig. \ref{fig:3},  we leverage a shared encoder \textit{$E_{\theta}(\cdot)$} to extract the features from each volumetric image in a sequence, which are later used for landmark detection. The 1D feature codes (for individual 3D volumes) are then processed by a following fully convolutional layer. The fusion layer, which further concatenates the codes of multiple input images, is expected to learn the sequential order (\textit{i.e.}, in phase index) of the inputs. That is, by employing a shuffle function \textit{$\mathcal{S}$} to randomly permutate the order of the input sequence, the network in Fig. \ref{fig:3} is trained to recover the correct phase index of the sequence, based on the features encoded through \textit{$E_{\theta}(\cdot)$}:
	\begin{equation}\label{eq:1}
	\begin{aligned}
	\mathcal{S}\{t1,t2,t3,t4,t5\}=E_{\theta}\bigl(\mathcal{S}(I_{t1},I_{t2},I_{t3},I_{t4},I_{t5})\bigr)
	\end{aligned}
	\end{equation}

	We thus formalize the above into a classification task as in Eq. \ref{eq:1}. The training is then supervised by the cross-entropy loss in classification. Through the above pre-training of the encoder, the encoder can \textcolor{black}{focus its} attention to the motion-affected region rather than the irrelative background in the encoding stage (\textit{e.g.}, the attention maps in Fig. \ref{fig:3}). 

	\subsubsection{Landmark Coordinate Extraction}\label{landmark}
	As shown in Fig. \ref{fig:4}, the input image first passes through the encoder and the decoder to generate $K$-channel outputs (note that the encoder is pre-trained already and will be further refined in the subsequent training). Then we apply a softmax layer in a spatial-wise manner to generate \textcolor{black}{$K$ heatmaps, \textit{i.e.,} $\mathcal{H}_{t}$. We then regress the landmark positions $L_{t}$ from the predicted heatmaps $\mathcal{H}_{t}$ as:} 
	\begin{equation}\label{eq:extract}
	\begin{aligned}
	L_{tk}=\Big(\sum_{x}^{\varOmega} (\mathcal{H}_{tk} \times \mathcal{G}_{x}), \sum_{y}^{\varOmega} (\mathcal{H}_{tk} \times \mathcal{G}_{y}), \sum_{z}^{\varOmega} (\mathcal{H}_{tk} \times \mathcal{G}_{z}) \Big), \\
	s.t. \sum\mathcal{H}_{tk}=1, \{k=1,...,K\}
	\end{aligned}
	\end{equation}
	where $(\mathcal{G}_{x},\mathcal{G}_{y},\mathcal{G}_{z})$ denotes the grid coordinate of the volumetric image as pre-mentioned $\mathcal{G}_{(x,y,z)}$ with the size of $(3, X, Y, Z)$. The grid coordinate $\mathcal{G}_{(x,y,z)}$ axis range (in $x, y, z$) is $[-1,1]$. We finally capture the exact coordinates of the extracted landmarks by localizing the position with the sum response in the heatmap, as in \cite{jakab2018unsupervised}.
	

	\subsubsection{Training Loss}
	We introduce four losses to assist in detecting the landmarks in 3D space. Among them, we firstly propose a landmark exclusive loss $\mathcal{L}_{1}$ to enhance the exploration in the 3D image space. Then we present a temporal coherence loss $\mathcal{L}_{2}$ to guide the landmarks to focus on the potential motion-affected region. Considering that the extracted landmarks should conform to human anatomy, we further propose a topological distribution loss $\mathcal{L}_{3}$. Finally, following \cite{siarohin2019first}, we leverage a consistency loss $\mathcal{L}_{4}$ to enhance the stability of landmark detection. More details are as below.
	
	The landmark detector is denoted as $L_{t}=D_{\zeta}(I_{t})$, where $D$ represents the detection network in volumetric space and $\zeta$ are the learnable parameters of the network. As shown in Fig. \ref{fig:4}, we use an end-to-end encoder-decoder architecture with skip-connections for generating two sets of heatmaps $\mathcal{H}_{t_{m}}$ and $\mathcal{H}_{t_{n}}$, separately, given two images $I_{t_{m}}$ and $I_{t_{n}}$ of \textcolor{black}{different phases}. We can derive the spatial coordinates $L_{t_{m}}$ and $L_{t_{n}}$ of the detected landmarks from the heatmaps $\mathcal{H}_{t_{m}}$ and $\mathcal{H}_{t_{n}}$.

	\textbf{Landmark Exclusive Loss} --- In order to increase the space exploration capability of landmark detection to enhance the richness of their spatial description, we assume there is mutual repulsion between any two landmarks if they are spatially in close proximity. We particularly apply a Gaussian distribution in volumetric space to describe each landmark centered on its own position as below,
	\begin{equation}\label{eq:2}
	\begin{aligned}
	\mathcal{F}_{tk}=\exp\Big ( -\dfrac{(L_{tk}-\mathcal{G}_{(x,y,z)})^{2}}{2\sigma^{2}} \Big ), (x,y,z) \in \varOmega
	\end{aligned}
	\end{equation}
	\textcolor{black}{where $\sigma$ is used to control the spread of each landmark, to tackle the uncertainty in its representation. $\mathcal{F}_{tk}$ is 3D Gaussian heatmap representing the landmark coordinate position. Note that \textcolor{black}{the value range} in each landmark's heatmap is \textcolor{black}{$(0,1]$}. The sum of two landmarks' heatmap will exceed 1 in certain coordinate if the two landmarks are close to each other, which should be penalized. The loss of the repulsive force is thus \textcolor{black}{defined} as:}


	\begin{equation}\label{eq:3}
	\begin{aligned}
	\mathcal{L}_{1}=  ReLU \bigg (\sum_{k=1}^{K} 
	\mathcal{F}_{tk} - 1 \bigg ) 
	\end{aligned}
	\end{equation}
	\textcolor{black}{By minimizing $\mathcal{L}_{1}$, we can avoid the case that all landmarks conglomerate in landmark detection.}
	
	
	\textbf{Temporal Coherence Loss} --- While the above loss drives all landmarks to explore the entire 3D space, we add temporal coherence loss to help the landmark detector better focus on the motion-affected region. Specifically, the extracted landmarks are expected to cover the motion-affected region, rather than the static background. Therefore, if two corresponding landmarks in the temporal sequence are static across different time-points, such landmarks will be penalized to be detected. Thus, given two corresponding landmarks $L_{t_{m}k}$ and $L_{t_{n}k}$, the self-supervised temporal coherence loss is defined as below:

	\begin{equation}\label{eq:4}
	\begin{aligned}
	\mathcal{L}_{2}=\sum_{k=1}^{K} \left\|
	\mathcal{F}_{t_{m}k} \times \mathcal{F}_{t_{n}k}\right\|_{1}
	\end{aligned}
	\end{equation}
	\textcolor{black}{where \textit{$\times$} is the element-wise product.}

	\textbf{Topological Distribution Loss} --- Considering that the extracted landmarks should largely conform to human anatomy, therefore, we apply a reference point distribution loss to the landmark detector.

	\begin{figure}[htbp]
		\begin{center}
			\includegraphics[width=1\linewidth]{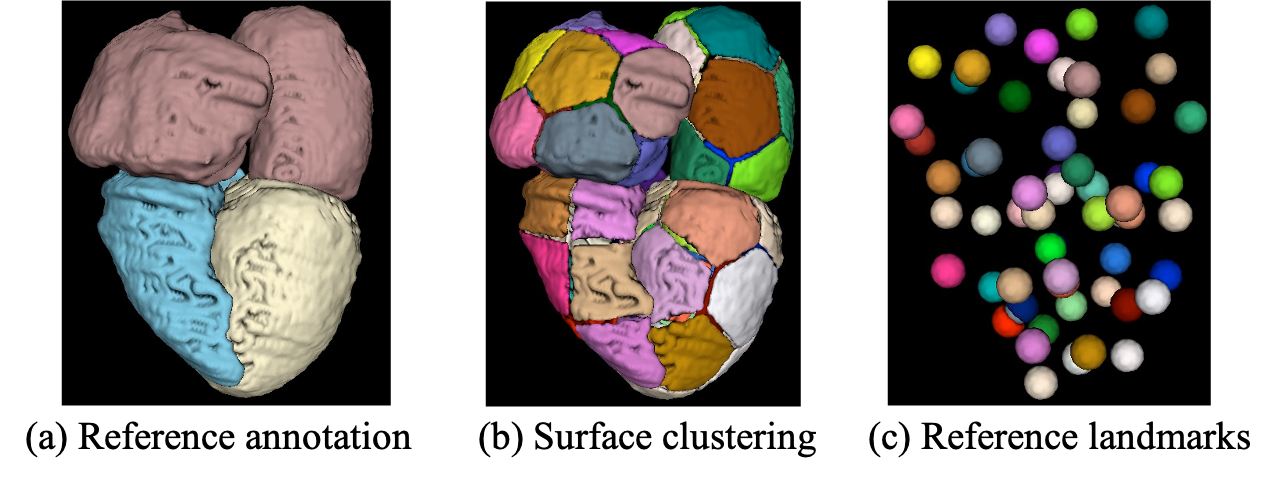}
		\end{center}
		\caption{Reference point distribution: (a) surface visualization of the reference annotations (from an atlas); (b) subgroup annotations with different colors after k-means clustering; (c) Reference points are derived as the center points of the clusters.} 
		\label{fig:5}
	\end{figure}

	\begin{figure*}[htbp]
		\centering
		\subfigure[Topological representation.]{
			\begin{minipage}[t]{0.5\linewidth}
				\centering
				\includegraphics[width=1\linewidth]{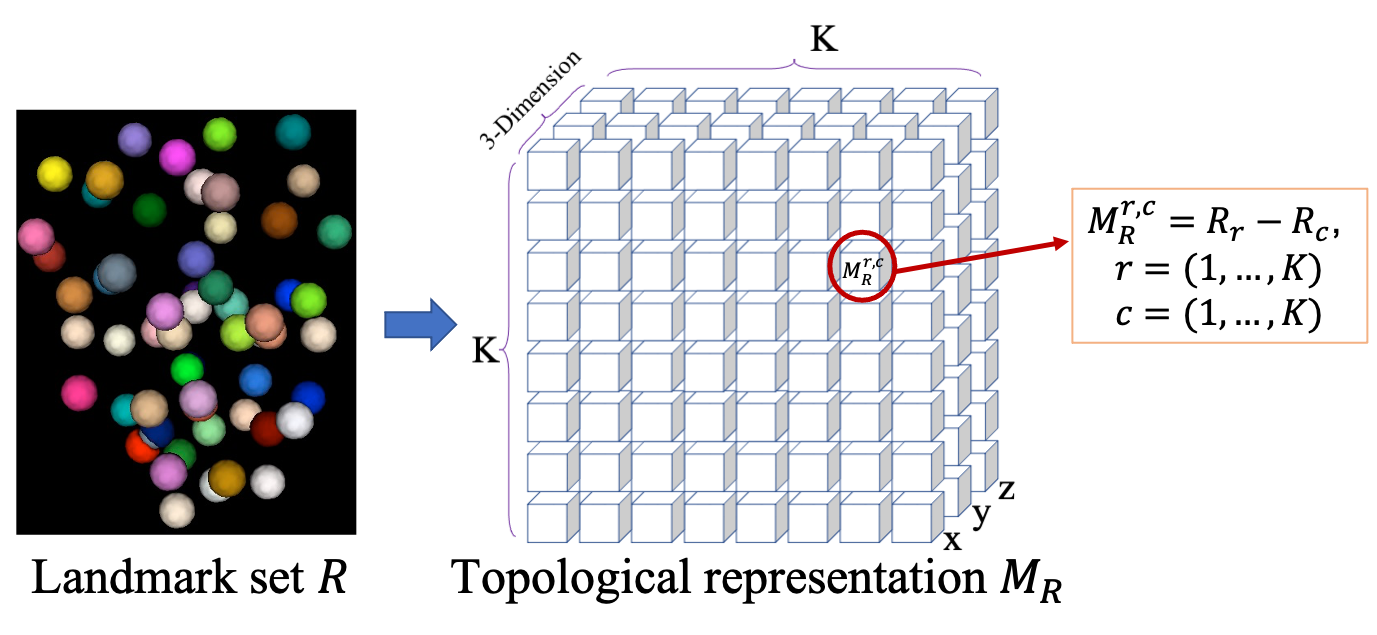}
				\label{fig:matrix}
			\end{minipage}%
		}%
		\subfigure[The optimal solution for topological distribution alignment.]{
			\begin{minipage}[t]{0.5\linewidth}
				\centering
				\includegraphics[width=1\linewidth]{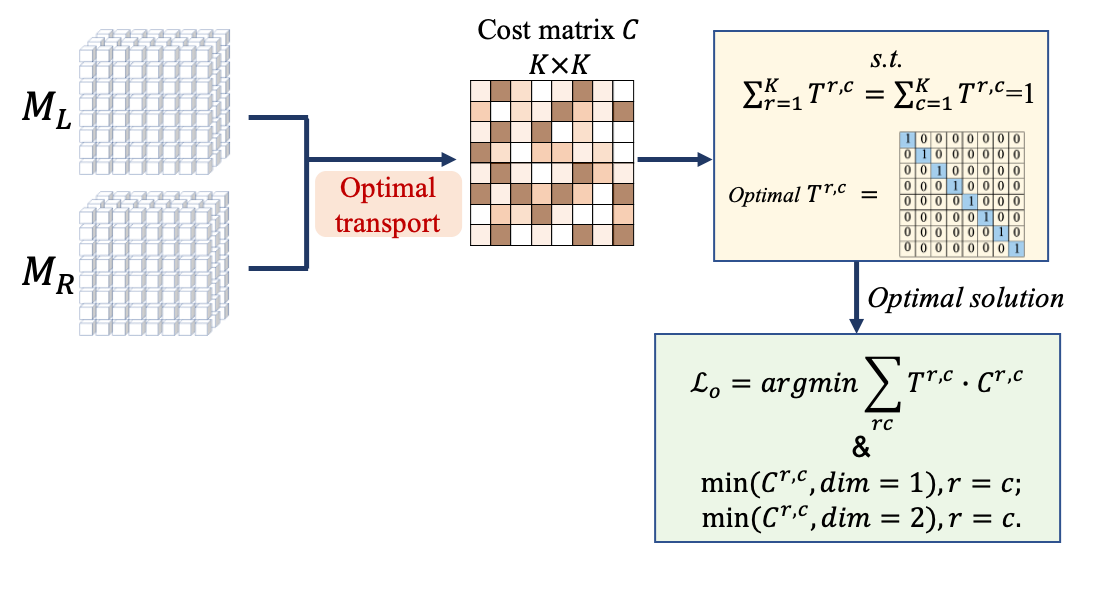}
				\label{fig:OT_s}
			\end{minipage}%
		}%
		\centering
		\caption{The optimal transport theory based topological loss: (a) indicates the calculation of the reference topological representation $M_{R}$ from its corresponding landmark set $R$; (b) depicts the optimal solution to align our detected landmarks to reference landmarks by minimizing the cost matrix $C$ (maximizing the correspondence) between the topological representation $M_{R}$ and $M_{L}$.}
		\label{fig:ot}
	\end{figure*}

	Fig. \ref{fig:5} depicts the process to extract the reference landmarks. We first choose an atlas downloaded from the public cardiac segmentation dataset \footnote{https://zmiclab.github.io/projects/mmwhs}. It is worthy to note that we only utilize a single atlas and the annotation, which can sufficiently establish reference point distribution in our work. Then, we extract the surface from the annotated segmentation as shown in Fig. \ref{fig:5}(a). The extracted surface represents the structural topology of the target organ. Next, we use k-means \cite{shi2011adaptive} to cluster the points on the surfaces based on their 3D coordinates, which produces several subgroups as shown in Fig. \ref{fig:5}(b). Finally, the center of each subgroup is used as the reference point distribution to represents the anatomical topology of the target organ as a whole (see Fig. \ref{fig:5}(c)).

	We then utilize the optimal transport theory \cite{villani2008optimal,kolkin2019style} to enforce the reference point distribution in landmark detection. Let $R_{k} (k = 1,...,K)$ denote the coordinates of the reference landmarks in the set. Then, the reference landmarks and our detected landmarks $L$ can be used to derive two tensors $\mathop{M_{R}}$ and $\mathop{M_{L}}$, both of which are sized $K\times K\times 3$ (see Fig. \ref{fig:matrix}), following Eq. \ref{eq:dis}
	\begin{equation}\label{eq:dis}
	\mathop{M_{L}} = L_{r} - L_{c}, \mathop{M_{R}} = R_{r} - R_{c}
	\end{equation}
	where $r$ and $c$ indicate the row and column indices of the tensors, respectively. Note that $M_{R}$ and $M_{L}$ encode mutual distances between landmarks in the reference and the detected sets. Each row/column of the two tensors can be regarded as representation for the specific landmark.
	
	Since the landmarks in the two sets $R$ and $L$ are presumably corresponding to each other, we can get high correlation between $M_{R}$ and $M_{L}$, given their one-to-one correspondence assignment. That is, we formulate the following optimal transport problem: 
	
	\begin{equation}\label{eq:6}
	\begin{aligned}
	\mathcal{L}_{3} = \mathop{argmin} \sum_{r}^{K}\sum_{c}^{K}T^{r,c}\times C^{r,c}, \\
	s.t. \sum\limits_{r}T^{r,c}=\sum\limits_{c}T^{r,c}=1
	\end{aligned}
	\end{equation}
	where \textit{$T$} is the binary transport matrix to establish one-to-one correspondences from the detected landmark set \textit{$L$} to the reference set \textit{$R$}. \textit{$C$} gauges the distances between the landmarks in the two sets. Thus, the element-wise product between \textit{$T$} and \textit{$C$}, as in Eq. \ref{eq:6}, penalizes two landmarks, if they are corresponding with each other across the two sets (thus activated by \textit{$T$}) yet differ a lot (with high distance in \textit{$C$}).
	
	
	In our implementation, we refer to the representation of the landmarks in \textit{$M_{L}$} and \textit{$M_{R}$} to derive the matrix \textit{$C$}. Specifically, the value at the $r$-th row and $c$-th column in \textit{$C$} is calculated as
	\begin{equation}\label{eq:c}
	\begin{aligned}
	C^{r,c} =  - \sum_{k=1}^{K}<M_{L}^{r,k},M_{R}^{c,k}>
	\end{aligned}
	\end{equation}
	

	
	Additionally, in our study, we expect the detected landmark set and the reference landmark set to have one-to-one correspondences. Thus, \textit{$T$} can be simplified as an identity matrix. Consequently, all entries of the optimal transport matrix \textit{$T$} will be 0 when \textit{$r\neq c$}, and 1 when \textit{$r=c$}. The above process is illustrated in Fig. \ref{fig:ot}.

	\textbf{Consistency Loss} --- To further improve the stability of landmark detection, we leverage a consistency loss inspired by \cite{siarohin2019animating}. For the landmarks extracted from each image, we can randomly generate an affine transformation matrix $\mathcal{A}=[ \mathcal{R}, \mathcal{B}] + [\mathcal{C}]$, where $\mathcal{R}$ denotes the rotation, $\mathcal{B}$ denotes the translation, and $\mathcal{C}$ is Gaussian noise. The generated affine transformation can warp image $I_{t_{m}}$ by $\mathcal{W}$ (cf., the warp layer in Fig. \ref{fig:2}) to become a \textit{new} image $\widetilde{I}_{t_{m}}$. And then the deformed image $\widetilde{I}_{t_{m}}$ can serve as a \textit{new} image for landmark detection, from which we extract another landmark set $\widetilde{L}_{t_{m}}=D_{\zeta}(\mathcal{W}(I_{t_{m}}\mid \mathcal{A}))$. The two landmark sets, given known affine transformation matrix $\mathcal{A}$, should be consistent following the loss in Eq. \ref{eq:5}:
	
	
	\begin{equation}\label{eq:5}
	\begin{aligned}
	\mathcal{L}_{4}=\sum_{k=1}^{K} \|\mathcal{W}(L_{tk}\mid \mathcal{A}) - \widetilde{L}_{tk}\|_{1}
	\end{aligned}
	\end{equation}

	\subsection{From Sparse Motion to Dense Motion - Motion Reconstruction Network}\label{motion}
	In this section, the estimation of the dense motion field $\varphi_{t_{m}\leftarrow t_{n}}$ from the detected landmarks ($L_{t_{m}},L_{t_{n}}$) and their sparse correspondences are introduced. We firstly compute the sparse displacement $V_{t_{m}k\leftarrow t_{n}k}, \{k = 1, ..., K\}$ from the detected landmarks $L_{t_{m}},L_{t_{n}}$. Then, we utilize a dense motion reconstruction network $\mathcal{M}$ (standard U-Net architecture) to estimate the dense motion field from the displacement of the detected landmarks. 
	
	
	The sparse displacement $V_{t_{m}k\leftarrow t_{n}k}, \{k = 1, ..., K\}$ links two temporally-consecutive landmarks $L_{t_{n}k}$ and $L_{t_{m}k}$ in $I_{t_{n}}$ and $I_{t_{m}}$, respectively. As the displacement $V_{t_{m}k\leftarrow t_{n}k}$ is spatially sparse and cannot directly input to our motion network $\mathcal{M}$, we first filling the entire $\widetilde{\varphi}_{t_{m}k\leftarrow t_{n}k}$ grid with the value of $V_{t_{m}k\leftarrow t_{n}k}$ and acquire a dense volumetric deformation field $\widetilde{\varphi}_{t_{m}k\leftarrow t_{n}k}$. We then use $\widetilde{\varphi}_{t_{m}k\leftarrow t_{n}k}$ to translate the image $I_{t_{n}}$ via the warp layer $\mathcal{W}$, which results in $K$ different deformed images $\widetilde{I}_{t_{m}k\leftarrow t_{n}k}$ in total. Thus, the deformed images $\widetilde{I}_{t_{m}k\leftarrow t_{n}k}$ are input to $\mathcal{M}$.
	
	
	However, the deformed \textit{$\widetilde{I}_{t_{m}k\leftarrow t_{n}k}$} only contains displacement information in images yet the landmark positions are missing. We thus further inject \textcolor{black}{the corresponding landmarks' heatmaps} into \textit{$\mathcal{F}$}. We specifically derive the variation between the corresponding heatmaps as
	\begin{equation}\label{eq:14}
	\begin{aligned}
	\mathcal{F}_{t_{m}k\leftarrow t_{n}k}= \mathcal{F}_{t_{m}k} - \mathcal{F}_{t_{n}k}
	\end{aligned}
	\end{equation}

	Finally, we concatenate the deformed images \textit{$\widetilde{I}_{t_{m}k\leftarrow t_{n}k}$} and the variation of heatmaps \textit{$\mathcal{F}_{t_{m}k\leftarrow t_{n}k}$} along the channel dimensionality, and feed them into the motion reconstruction network \textit{$\mathcal{M}$} to fuse the \textit{$\widetilde{\varphi}_{t_{m}k\leftarrow t_{n}k}$} into final dense motion field \textit{$\varphi_{t_{m} \leftarrow t_{n}}$} as 
	\begin{equation}\label{eq:M}
	\begin{aligned}
	\varphi_{t_{m} \leftarrow t_{n}}= \sum_{k=1}^{K}\mathcal{M}(\widetilde{I}_{t_{m}k\leftarrow t_{n}k},\mathcal{F}_{t_{m}k\leftarrow t_{n}k}) * \widetilde{\varphi}_{t_{m}k\leftarrow t_{n}k}
	\end{aligned}
	\end{equation} 
	
	The training of \textit{$\mathcal{M}$} is supervised by a voxel-wise similarity loss calculated from the intensity image and its corresponding heatmaps as
	\begin{equation}\label{eq:13}
	\begin{aligned}
	\mathcal{L}_{d}=& \parallel \mathcal{W}(\mathcal{M}_{\gamma}(\varphi_{t_{m}\leftarrow t_{n}}),I_{t_{n}}) - I_{t_{m}} \parallel_{2} + \\
	&\sum_{k=1}^{K}\parallel \mathcal{W}(\mathcal{M}_{\gamma}(\varphi_{t_{m}\leftarrow t_{n}}),\mathcal{H}_{t_{n}k}) * \mathcal{G}_{x,y,z} - L_{t_{m}k} \parallel_{1} 
	\end{aligned}
	\end{equation}
	

	~\\
	
	In overall, by integrating the landmark detection network and the dense motion reconstruction network, our final training loss \textit{$\mathcal{L}$} is defined as
	
	\begin{equation}\label{eq:15}
	\begin{aligned}
	\mathcal{L}= \lambda_{1}\mathcal{L}_{1} + \lambda_{2}\mathcal{L}_{2} + \lambda_{3}\mathcal{L}_{3} + \lambda_{4}\mathcal{L}_{4} + \lambda_{d}\mathcal{L}_{d}
	\end{aligned}
	\end{equation}
	
	The weights \textit{$\lambda_{1}=1, \lambda_{2}=0.5, \lambda_{3}=20, \lambda_{4}=100, \lambda_{d}=500$} have been set empirically. As shown in Fig. \ref{fig:2}, we thus achieve the \textit{dense-sparse-dense} transferring and estimate the motion field between the input images. In the final, the reconstructed motion field can be employed as an initial transformation field and further refined via existing registration methods \cite{wang2015predict}. The refinement aims to boost the quality in estimating the motion field in detail, while the initial transformation produced by DSD has already addressed large deformation. Thus, the refinement only needs a small amount of iterative computation, which is highly efficient.
	
	
	\section{Experiments}
	
	\subsection{Materials and Implementation details}
	
	We demonstrate our method with three datasets: 4D Cardiac CT (\textit{4D-C-CT}), 4D Lung CT (\textit{4D-L-CT}) \cite{hugo2016data}, and \textcolor{black}{4D MR cardiac ACDC \cite{bernard2018deep}}. 
	
	\begin{figure}[!htbp]
		\centering
		\subfigure[4D-C-CT.]{
			\begin{minipage}[t]{0.33\linewidth}
				\centering
				\includegraphics[width=1\linewidth]{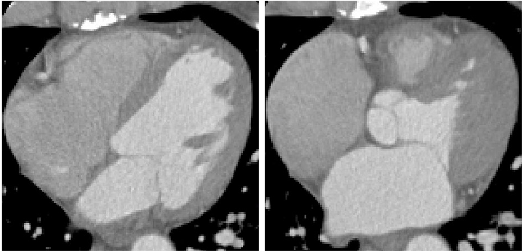}
				\label{fig:heart}
			\end{minipage}%
		}%
		\subfigure[4D-L-CT.]{
			\begin{minipage}[t]{0.33\linewidth}
				\centering
				\includegraphics[width=1\linewidth]{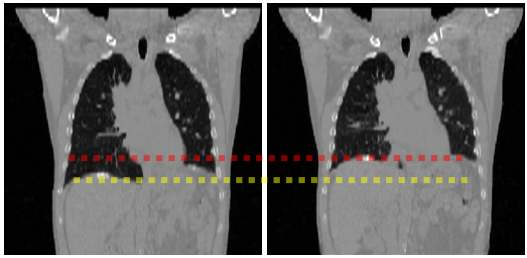}
				\label{fig:lung}
			\end{minipage}%
		}%
		\subfigure[ACDC.]{
			\begin{minipage}[t]{0.33\linewidth}
				\centering
				\includegraphics[width=1\linewidth]{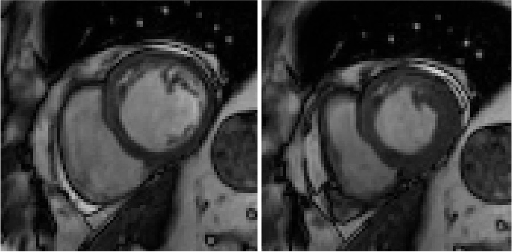}
				\label{fig:acdc}
			\end{minipage}%
		}%
		\centering
		\caption{A snapshot of the selected data at two time-points with the largest motion difference: (a) is the \textit{4D-C-CT} (axial views) showing the cardiac phases in the end-diastole (ED) and end-systole (ES), respectively of left and right; (b) is the \textit{4D-L-CT} (sagittal views) depicting the lungs in the states of maximum breath holding and minimum exhalation. The red line and yellow lines highlight the volume changes of the lung between two phases. \textcolor{black}{(c) is the \textit{ACDC} depicting the cardiac MR images in the ED and ES, respectively of left and right.}}
		\label{fig:6}
	\end{figure}

	\subsubsection{4D-C-CT} Fig. \ref{fig:heart} shows a snapshot of randomly sampled cardiac volume slices from a sequence. The \textit{4D-C-CT} dataset consists of 18 subjects, each having 5 time-points (image volumes) capturing half cardiac cycle from ED to ES. Since the dataset contains several patients with left ventricular aneurysm resulting in abnormally subtle deformation, we select 12 patient data with large volume changes (volume changing rate from 0.37 to 0.75), and only test on these selected challenging datasets. Each volume is characterized by a high intra-slice (x- and y-) resolution ranging from 0.32 to 0.45mm and inter-slice (z-) resolution from 0.37 to 0.82mm. 
	
	\subsubsection{4D-L-CT} Fig. \ref{fig:lung} shows a snapshot of lung volume slices from a sequence. The \textit{4D-L-CT} dataset consists of 20 patients, each having 10 time-points representing the whole respiratory cycle. 4D-L-CT images were acquired using a 16$-$slice, helical CT scanner (Brilliance Big Bore, Philips Medical Systems) with a slice thickness of 3 mm and $512 \times512$ axial resolution ($\sim1$ mm pixel size). 
	
	\textcolor{black}{\subsubsection{ACDC} Fig. \ref{fig:acdc} shows a snapshot of cardiac MR middle slices from a sequence. The ACDC dataset consists of 100 patients, each having multiple time-points representing the whole cardiac cycle. It has lower imaging resolution - the in-plane voxel spacing ranges from 1.37 to 1.68 mm and the inter-slice spacing ranges from 5 to 10mm. Note that, as inter-slice spacing is large in \textit{ACDC}, we have to change all the comparisons to 2D, and evaluate on three middle slices in each volume.}

	\begin{table*}[htbp]\scriptsize
		\centering
		\color{black}\caption{\textcolor{black}{Motion estimation results on the \textit{4D-C-CT} dataset among the individual time-points, from ED to ES for all the comparison methods.}}
		\setlength{\tabcolsep}{.5mm}{
			\begin{tabular}{c|cccc|cccc|cccc|cccc|cccc}
				\hline
				{} &  \multicolumn{4}{c|}{Interval-1} & \multicolumn{4}{c|}{Interval-2} & \multicolumn{4}{c|}{Interval-3} & \multicolumn{4}{c|}{Interval-4}& \multicolumn{4}{c}{\textit{Mean}}\\
				\hline
				{}   & DSC   & HD  & ASD & RVD  & DSC   & HD  & ASD & RVD & DSC   & HD& ASD & RVD& DSC   & HD& ASD & RVD& DSC   & HD& ASD & RVD\\
				\hline
				FFD & 0.972 & 4.45 & 0.414 & 0.011 & 0.943 & 6.88 & 0.759 & 0.028 & 0.884 & 7.21 & 1.406 & 0.086 & 0.838& 8.65 & 1.935 & 0.172 & 0.909$\pm$0.062&6.80$\pm$1.95 & 1.13$\pm$0.75 & \textbf{0.074$\pm$0.100}\\
				Demons  & 0.977 & 3.82 & 0.335 & 0.006 & 0.958 & 4.64 & 0.554 & 0.027 & 0.888 & 5.69 & 1.419 & 0.199  & 0.825& 7.37 & 2.164 & 0.377 & 0.912$\pm$0.068& 5.38$\pm$1.87 & 1.12$\pm$0.96 & 0.152$\pm$0.212\\
				FeaturePoints  & 0.943 & 4.63 & 0.709 & 0.041 & 0.867 & 10.03 & 1.454 & 0.237 & 0.736 & 11.22 & 2.895 & 0.699 & 0.650& 12.96 & 3.970 & 1.071 & 0.800$\pm$0.150& 9.72$\pm$3.43 & 2.26$\pm$1.61 & 0.512$\pm$0.607\\
				VoxelMorph & 0.951 & 4.76 & 0.463 & 0.017 & 0.905 & 6.12 & 1.135 & 0.122 & 0.836 & 7.57 & 2.066 & 0.203  & 0.761 & 9.02 & 2.991 & 0.511 & 0.863$\pm$0.090 &6.87$\pm$2.30 & 1.64$\pm$0.99 & 0.213$\pm$0.226\\
				VoxelMorph + Demons & 0.976 & 3.34 & 0.347 & 0.008 & 0.960 & 3.96 & 0.547 & 0.033 & 0.891 & 6.36 & 1.226 & 0.162 & 0.820 & 7.19 & 1.761 & 0.332 & 0.911$\pm$0.071 &5.21$\pm$1.89 & 0.97$\pm$0.88 & 0.134$\pm$0.207\\
				DSD (Ours) & 0.954 & 4.09 & 0.510 & 0.022 & 0.929 & 4.96 & 0.981 & 0.093 & 0.853 & 7.14 & 2.115 & 0.208 & 0.792 & 8.41 & 2.939 & 0.447 & 0.882$\pm$0.084 & 6.16$\pm$2.18 & 1.66$\pm$0.92 & 0.192$\pm$0.258 \\
				DSD + Demons (Ours) & 0.978 & 3.25 & 0.266 & 0.005 & 0.964 & 3.90 & 0.427 & 0.023 & 0.916 & 5.70 & 1.066 & 0.117 & 0.862 & 6.22 & 1.440 & 0.254 & \textbf{0.930$\pm$0.058} & \textbf{4.81$\pm$1.56} & \textbf{0.80$\pm$0.72} & 0.100$\pm$0.184 \\
				\hline
		\end{tabular}}
		\label{table:3}
	\end{table*}
	
	\begin{figure*}[htbp]
		\begin{center}
			\includegraphics[width=1\linewidth]{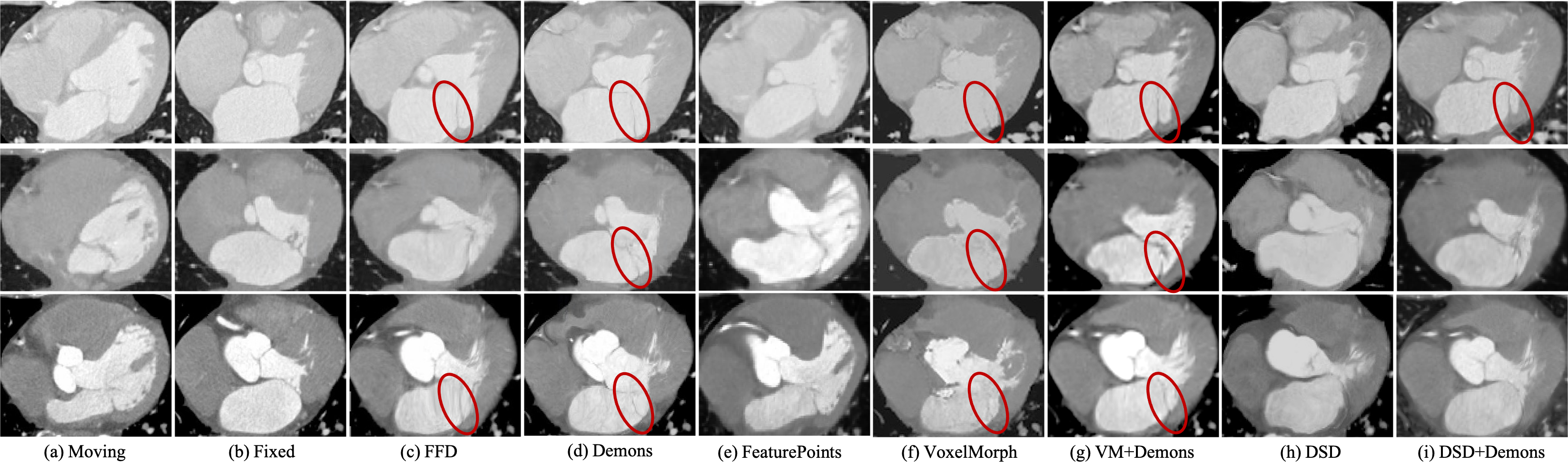}
		\end{center}
		\caption{Visual results of three samples from \textit{4D-C-CT}. (a) and (b) indicate the input paired images. (c) - (h) indicate the deformed moving images estimated from the comparison methods. The red circle describes the region associating with folding artifacts and error estimations.}
		\label{fig:8}
	\end{figure*}
	
	
	We apply contrast-normalization to all the images, consistent with other similar experiments \cite{jang2017automatic}. All the networks are implemented using Pytorch library and trained on two 11GB Nvidia 1080Ti GPUs. The landmark detection model is trained with a learning rate of \textit{$2\times 10^{-5}$} while the dense motion reconstruction network is trained with a learning rate of \textit{$6\times 10^{-5}$}. In the evaluations of 4D-C-CT and 4D-L-CT, we use four-fold cross-validation on both datasets. \textcolor{black}{In the evaluation of ACDC dataset, we random select 70 training / 30 testing subjects.} 

	\subsection{Evaluation and Metrics}
	
	We conduct comprehensive experiments to validate our proposed method - \textit{DSD} (without subsequent refinement) and \textit{DSD + Demons} (with refinement by Demons). \textcolor{black}{\textit{\textbf{First}}, we compare our proposed method with three different categories of the state-of-the-art motion estimation methods, including conventional non-rigid image registration method, unsupervised CNN-based registration, and feature point based registration. We implement all the compared methods following their default parameter settings in the same environment, including the same training set, and same testing set:}
	\textcolor{black}{\begin{enumerate}
			\item Conventional non-rigid image registration methods - \textit{Demons} \cite{vercauteren2009diffeomorphic} and Free-Form Deformation (\textit{FFD}) \cite{modat2010fast}. Demons utilizes the intensity similarity of the images to derive the driving force, which deforms the moving image to the fixed image. FFD method requires to initialize a group of control points, and optimizes the control points’ displacement to interpolate the deformation field, which warps the moving image to the fixed image.
			\item Unsupervised deep CNN-based method – \textit{VoxelMorph} \cite{balakrishnan2019voxelmorph}. The deep CNN network is trained via the loss from the similarity of image appearance and the smoothness of the deformation field in an unsupervised manner. In the inference stage, it can directly yield the entire deformation field given the moving and the fixed images.
			\item Feature point based method - FeaturePoints \cite{hosseini2021non}. Image local feature points are firstly extracted via the popular scale-invariant feature transform (SIFT) descriptors, and then utilized to register the two images.
	\end{enumerate}}
	
	
	\textcolor{black}{\textit{\textbf{Next}}, for unsupervised landmark detection, we compare it with the state-of-the-art first-order motion model (FOMM) \cite{siarohin2019first}\footnote{https://aliaksandrsiarohin.github.io/first-order-model-website/} in the ablation study. Note that, since FOMM is originally presented for 2D images, we have adapted FOMM to 3D and regarded 3D-FOMM as the baseline for comparison.}
	
	\textcolor{black}{For all comparisons, we implement them using their default parameter settings. For Demons and FFD, we implement them via SimpleITK library.} For evaluation, we use the popular metrics including Hausdorff Distance (HD), Dice Similarity Coefficient (DSC), \textcolor{black}{Average Surface Distance (ASD), and Relative volume difference (RVD). Lower HD, ASD, RVD} and higher DSC indicate better performance.

	\section{Results and Discussion}
	


	\subsection{Comparison with the state-of-the-art motion estimation methods}
	
	We have applied our \textit{DSD} framework to both 4D-C-CT and 4D-L-CT datasets for the evaluation of its performance. Furthermore, we have employed the motion field produced by deep-learning-based methods as an initialization, and then refine it by Demons, following the strategy in \cite{wang2015predict}. Note that, the FFD is not selected to refine our motion field due to it does not support the initial deformation setting.

	\subsubsection{4D-C-CT dataset}

	The quantitative comparison for cardiac motion estimation is shown in Table \ref{table:3}. As the motion estimation method should be able to handle the time-points with varying intervals, we also evaluate the performance with different time intervals from ED to ES in Table \ref{table:3}. It shows that all the methods have obtained relatively high performances when the time interval is small (\textit{i.e.}, implying small deformation), but the performances drop as the time interval becomes larger (\textit{i.e.}, implying large deformation). This is because of the fact that large deformation causes the accuracy of motion estimation to decrease. In contrast, our \textit{DSD + Demons} exhibits the improvement in motion estimation accuracy relative to the comparison methods, and the improvement increases as the time interval become larger. We attribute this to the motion estimation guided by the extracted landmarks that can effectively handle the large deformation by avoiding the error estimation. 
	
	As highlighted in Fig. \ref{fig:8}, where \textit{FFD} and \textit{Demons} are prone to errors for the images with large motion, which may result in folding in image deformation. The deep-learning-based \textit{VoxelMorph} slightly alleviates the error deformation, but there are problems in underestimating the deformation as in Fig. \ref{fig:8}. In contrast to these methods, our proposed method first extracts landmarks to represent the topological structure of the target organ and estimates the dense motion field by considering the dynamic changes of the organ's morphology. While our \textit{DSD} exhibits the great capability of \textcolor{black}{modeling} the large motion from detected landmarks, the local appearance must be considered for subtle refinement of the motion field.

	\begin{table*}[htbp]\scriptsize
		\centering
		\color{black}\caption{\textcolor{black}{Motion estimation results on the \textit{4D-L-CT} dataset among the individual time-points, from maximum breath holding to minimum exhalation for all the comparison methods.}}
		\setlength{\tabcolsep}{.5mm}{
			\begin{tabular}{c|cccc|cccc|cccc|cccc|cccc}
				\hline
				{} &  \multicolumn{4}{c|}{Interval-1} & \multicolumn{4}{c|}{Interval-2} & \multicolumn{4}{c|}{Interval-3} & \multicolumn{4}{c|}{Interval-4}& \multicolumn{4}{c}{\textit{Mean}}\\
				\hline
				{}   & DSC   & HD  &  ASD  &  RVD   & DSC   & HD  &  ASD  &  RVD  & DSC   & HD &  ASD  &  RVD& DSC  & HD &  ASD  &  RVD& DSC   & HD &  ASD  &  RVD\\
				\hline
				FFD & 0.985 & 4.70 & 0.115 & 0.003 & 0.981 & 4.63 & 0.155 & 0.007 & 0.977 & 4.75 & 0.186 & 0.009 & 0.975& 5.34 & 0.206 & 0.010 & 0.979$\pm$0.005& \textbf{4.85$\pm$0.578} & 0.17$\pm$0.06 & 0.008$\pm$0.005\\
				Demons  & 0.987 & 4.67 & 0.105 & 0.003 & 0.982 & 4.61 & 0.140 & 0.009 & 0.978 & 4.78 & 0.181 & 0.017 & 0.973 & 5.41 & 0.219 & 0.027 &0.980$\pm$0.005& 4.87$\pm$0.584 & 0.17$\pm$0.07 & 0.015$\pm$0.013\\
				FeaturePoints  & 0.937 & 4.28 & 0.508 & 0.012 & 0.931 & 5.19 & 0.521 & 0.032 & 0.925 & 7.62 & 0.562 & 0.048 & 0.918 & 9.07 & 0.594 & 0.069 &0.925$\pm$0.014& 7.18$\pm$2.178 & 0.56$\pm$0.08 & 0.048$\pm$0.033\\
				VoxelMorph & 0.979 & 4.98 & 0.068 & 0.006 & 0.973 & 5.05 & 0.113 & 0.010 & 0.970 & 5.16 & 0.130 & 0.015 &0.967 & 5.07 & 0.141 & 0.020 & 0.972$\pm$0.005&5.06$\pm$0.523 & \textbf{0.11$\pm$0.03} & 0.007$\pm$0.010\\
				VoxelMorph + Demons & 0.986 & 4.86 & 0.073 & 0.000 & 0.982 & 5.18 & 0.154 & 0.008 & 0.979 & 5.21 & 0.193 & 0.015 &0.975 & 5.43 & 0.209 & 0.024 & 0.981$\pm$0.005&5.17$\pm$0.626 & 0.17$\pm$0.05 & 0.013$\pm$0.009\\
				DSD (Ours) & 0.983 & 4.79 & 0.092 & 0.003 & 0.977 & 5.16 & 0.154 & 0.008 & 0.970 & 5.19 & 0.193 & 0.015  & 0.967& 5.34 & 0.209 & 0.024 & 0.974$\pm$0.006&5.12$\pm$0.651 & 0.17$\pm$0.05 & 0.013$\pm$0.009\\
				DSD + Demons (Ours) & 0.987 & 4.43 & 0.084 & 0.002 & 0.983 & 4.87 & 0.123 & 0.003 & 0.980 & 4.94 & 0.144 & 0.005 & 0.979& 5.17 & 0.164 & 0.006 & \textbf{0.982$\pm$0.005}&4.85$\pm$0.597 & 0.13$\pm$0.03 & \textbf{0.004$\pm$0.003}\\
				\hline
		\end{tabular}}
		\label{table:4}
	\end{table*}
	
	\begin{figure*}[htbp]
		\begin{center}
			\includegraphics[width=1\linewidth]{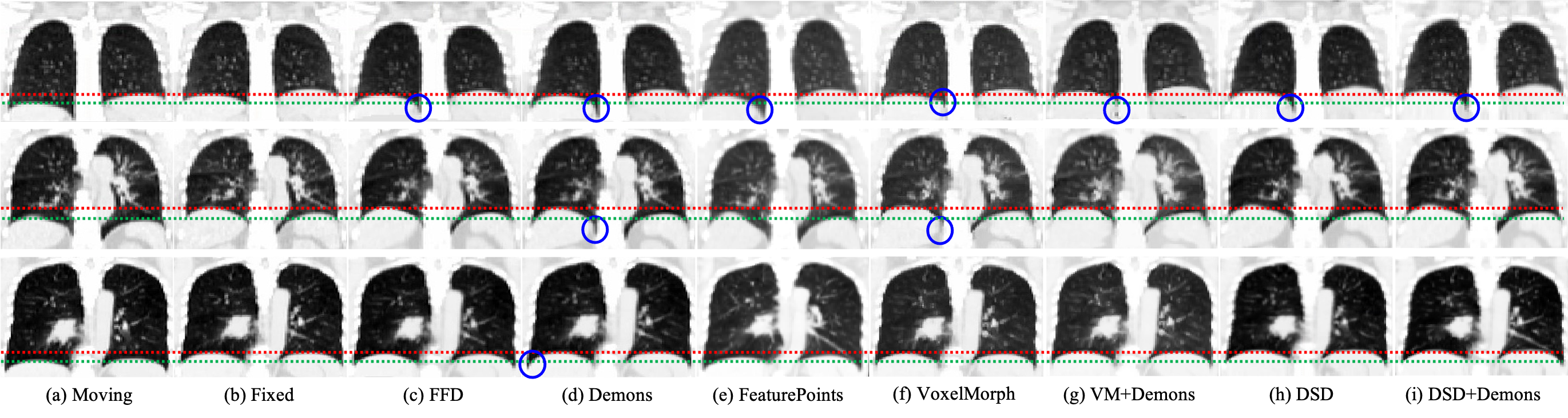}
		\end{center}
		\caption{Visual results of three samples from \textit{4D-L-CT}. The red line and green line describe the position of the right diaphragm (in anatomy) at the end of expiration and the end of inspiration respectively. The blue circle indicates the horizontal over-squeezing regions where should move upward.}
		\label{fig:9}
	\end{figure*} 
	
	Overall, our \textit{DSD + Demons} scheme achieves the best performance (\textit{DSC=0.930, HD=4.81}), and \textit{VoxelMorph + Demons} is ranked second-best (\textit{DSC=0.911, HD=5.21}). This reveals that our DSD framework can effectively guarantee the accuracy of motion estimation compared with VoxelMorph which is the state-of-the-art deep learning method. In addition, compared with the method only using conventional Demons (\textit{DSC=0.912, HD=5.38}), our \textit{DSD + Demons} also has an improvement in the accuracy of the estimations. We attribute this to the fact that the conventional methods, \textit{i.e.}, Demons and FFD, are greatly limited by their optimization based on local image similarity which leads to the inability to effectively estimate large deformation. 

	\subsubsection{4D-L-CT dataset}

	Table \ref{table:4} lists the quantitative motion estimation results on the \textit{4D-L-CT} dataset. Similar to the evaluation of cardiac data, we also compare the results with respect to different time intervals between the two input time-points. Our method \textit{DSD + Demons} achieves the best registration accuracy. In fact, the performance of all the compared methods is similar on the \textit{4D-L-CT} dataset with only 0.01 difference in DSC between the highest (\textit{DSD + Demons}: \textit{DSC=0.982, HD=4.85}) and the lowest (\textit{VoxelMorph}: \textit{DSC=0.972, HD=5.06}). This is because the anatomical structures of the lungs are relatively simple, and the motion is easy to estimate compared to the cardiac structure.

	Fig. \ref{fig:9} depicts the warped outputs for all the methods. The temporal changes of lungs in the appearance and shapes are generally mild between the phases of maximum breath holding and minimum exhalation (\textit{c.f.}, green lines and red lines in Fig. \ref{fig:9}). Although the conventional method \textit{Demons} achieves a well quantitative performance, there is unfortunately an over-squeezing in the deformation, and thus artifacts appear in the deformed image. For instance, as indicated by the blue circle in Fig. \ref{fig:9}, the estimated deformation on the lower lobes of the lungs is horizontally over-squeezed, where the correct deformation (see the fixed image and the moving image in Fig. \ref{fig:9}) should have been an upward movement. In contrast, our proposed method is able to effectively reduce the error of motion estimation in the stenosis of the lower lobe of the lung.

	\begin{table*}[htbp]
		\centering
		\color{black}\caption{\textcolor{black}{Motion estimation results on the ACDC dataset among the right ventricle (RV), left ventricle myocardium (LVM), and left ventricle (LV) between ED and ES}}
		\begin{tabular}{c|cccc|cccc|cccc}
			\hline
			{} &  \multicolumn{4}{c|}{RV} & \multicolumn{4}{c|}{LVM} & \multicolumn{4}{c}{LV}\\
			\hline
			{}   & DSC   & HD & ASD & RVD    & DSC   & HD  & ASD & RVD  & DSC   & HD& ASD & RVD \\
			\hline
			FFD & 0.771 & 7.324 & 3.202 & 0.647 & 0.734 & 6.364 & 1.414& 0.144 & 0.847& 3.695 & 1.593 & 0.266\\
			Demons  & 0.735 & 8.513 & 3.724 & 0.812 & 0.815 & 4.882 & 1.248& \textbf{0.101} & 0.829& 4.494 & 1.895 & 0.381\\
			FeaturePoints & 0.649 & 10.219 & 4.905 & 1.171 & 0.560 & 8.436 &2.241 & 0.137 & 0.632&8.325 & 4.548 & 1.247\\
			VoxelMorph & 0.771 & \textbf{3.991} & 2.439 & 0.188 & 0.731 & 4.321 &1.564 & 0.171 & 0.763&3.981&2.761&0.270\\
			DSD (Ours) & \textbf{0.806} & 4.240 & \textbf{1.647} & \textbf{0.246} & \textbf{0.842} & \textbf{3.672} & \textbf{1.242}& 0.116 & \textbf{0.876}&\textbf{3.073}&\textbf{1.323}&\textbf{0.234}\\
			\hline
		\end{tabular}
		\label{table:5}
	\end{table*}
	
	\begin{figure*}[htbp]
		\begin{center}
			\includegraphics[width=1\linewidth]{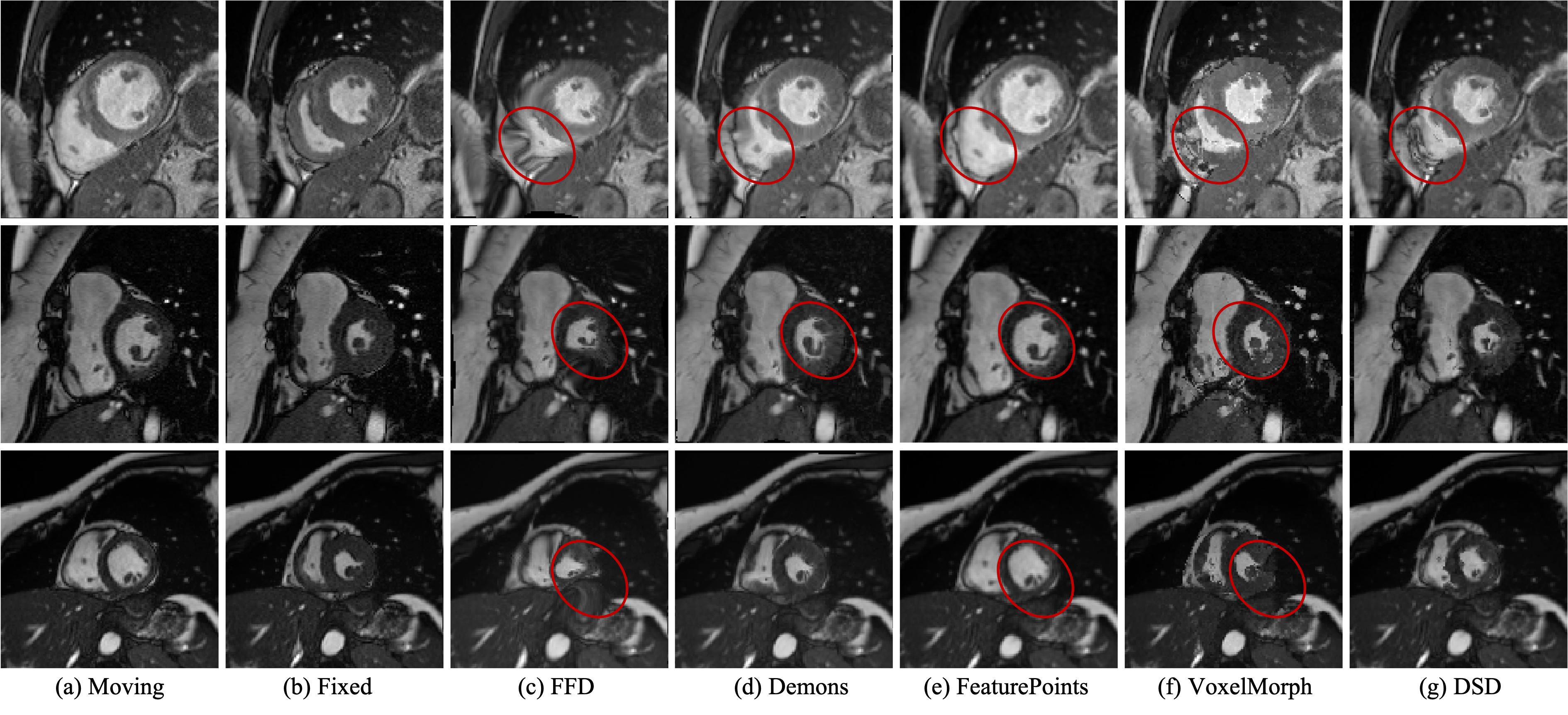}
		\end{center}
		\caption{\textcolor{black}{Visual results of three samples from the ACDC dataset: (a) and (b) indicate the input images; (c) - (f) indicate the deformed moving images by individual methods. The red circles highlight comparisons regarding folding artifacts and errors in the deformed images.}}
		\label{fig:10}
	\end{figure*} 
	
	\subsubsection{\textcolor{black}{ACDC dataset}}
	\textcolor{black}{Table \ref{table:5} lists the quantitative motion estimation results on the \textit{ACDC} dataset. Our proposed \textit{DSD} method achieves the outstanding performance across most evaluations. As shown in Fig. \ref{fig:10}, \textit{FFD} and \textit{Demons} are prone to errors in motion estimation. For feature point based method, the moving image is not effectively deformed, because the extracted feature points from the two time-point images lack consistency. Meanwhile, \textit{VoxelMorph} obtains relatively accurate estimation, but it may still result in folding in image deformation as shown in Fig. \ref{fig:10}. In contrast, our proposed DSD method is able to effectively and accurately track the motion in the images.} 

	\subsection{Ablation study of the DSD method}
	
	We also conduct an ablation study for our unsupervised landmark detection network, which is a critical component in our framework. \textcolor{black}{To effectively discuss the impact of our proposed losses on the accuracy of motion estimation and landmark detection, we analyze our proposed method with three groups of different combinations of loss functions following the scheme in \cite{huang2019knowledge}, i.e., Only-Keeping, Removing and Accumulating, as shown in Table \ref{table:1} shown. In the ‘Only-Keeping’ subgroup, we only keep one selected loss function to combine with the baseline model. For the second ‘Removing’ subgroup, we remove the selected loss function and employ all other proposed loss functions. In the last ‘Accumulating’ subgroup, we increasingly add the loss terms to the baseline model in the order of $\mathcal{L}_{1}$, $\mathcal{L}_{2}$ and $\mathcal{L}_{3}$. Note that, the baseline model (3D-FOMM) uses $\mathcal{L}_{4}$ and $\mathcal{L}_{d}$. We thus only investigate $\mathcal{L}_{1}$, $\mathcal{L}_{2}$ and $\mathcal{L}_{3}$ in this ablation experiment.}
	

	\begin{table*}[htbp]
		\centering
		\color{black}\caption{\textcolor{black}{The quantitative performance of spatiotemporal motion field estimation on the \textit{4D-C-CT} dataset for different ablation settings.}}
		\begin{tabular}{c|cccc|cccc|cccc}
			\hline
			{} &  \multicolumn{4}{c|}{Only-Keeping} & \multicolumn{4}{c|}{Removing} & \multicolumn{4}{c}{Accumulating}\\
			\hline
			{}   & DSC   & HD & ASD & RVD    & DSC   & HD  & ASD & RVD  & DSC   & HD& ASD & RVD \\
			\hline
			Baseline (3D-FOMM) & 0.783 & 8.26 & 3.36 & 0.604 & - & - & -& - & 0.783& 8.26 & 3.36 & 0.604 \\
			$\mathcal{L}_{1}$  & 0.837 & 7.46 & 2.43 & 0.356 & 0.807 & 8.56 & 2.78 & 0.518 &0.837 & 7.46 & 2.43 & 0.356 \\
			$\mathcal{L}_{2}$  & 0.806 & 8.34 & 2.92 & 0.462 & 0.880 & 6.42 & 1.51 & 0.196 & 0.842 & 7.19 & 2.34 & 0.261\\
			$\mathcal{L}_{3}$ & 0.799 & 8.72 & 2.76 & 0.508 & 0.842 & 7.19 & 2.34 & 0.261 & 0.882 &6.16&1.66&0.192\\
			\hline
		\end{tabular}
		\label{table:1}
	\end{table*}
	
\begin{figure}[!htbp]
		\begin{center}
			\includegraphics[width=1\linewidth]{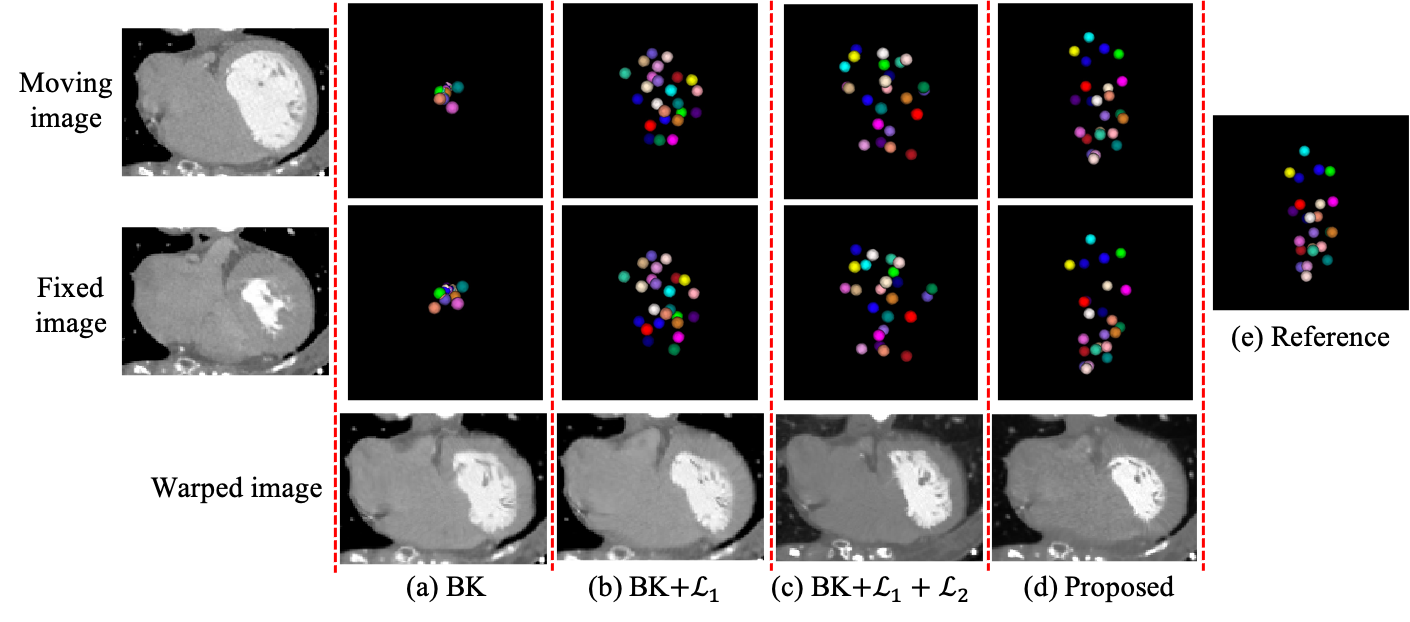}
		\end{center}
		\caption{Qualitative results of a selected sample from 4D-C-CT. It visually compares the extracted landmarks with different losses enabled.}
		\label{fig:heart_kp}
	\end{figure}

	\textcolor{black}{The motion estimation results in the ablation study are listed in Table \ref{table:1} for \textit{4D-C-CT}. Compared to the baseline model \textit{3D-FOMM}, our \textit{$\mathcal{L}_{1}$} loss greatly improves the effectiveness of landmark detection and also the accuracy in motion estimation. In contrast, using \textit{$\mathcal{L}_{2}$} and \textit{$\mathcal{L}_{3}$} alone does not significantly improve the model performance compared to the baseline, as shown in the 'Only-Keeping' subgroup in Table \ref{table:1}. However, only relying on \textit{$\mathcal{L}_{1}$} and \textit{$\mathcal{L}_{2}$} losses is not able to handle the paired images with large motion. The simultaneous use of \textit{$\mathcal{L}_{1}$} and \textit{$\mathcal{L}_{3}$} can further significantly improve the accuracy of motion estimation. We attribute this to that our specific designed topological loss $\mathcal{L}_{3}$ effectively establish the conversation between the detected landmarks and the target organ's anatomy. Therefore, the detected landmarks accurately reflect the dynamic changes of the target organ which can be further used to guide the motion estimation. Compared with $\mathcal{L}_{1}$ and $\mathcal{L}_{3}$, $\mathcal{L}_{2}$ does not greatly improve the accuracy of the model. We attribute this to that its effect mainly manifested in improving the description of landmark representation, especially without $\mathcal{L}_{3}$, as shown in Fig. \ref{fig:heart_kp}. Meanwhile, $\mathcal{L}_{d}$ and subsequent added $\mathcal{L}_{3}$ will weaken its effect on the model performance because they also can guide model to pay attention to the motion-affected region.}
	
	
	Fig. \ref{fig:heart_kp} presents the landmark distributions with different loss settings. Adding the driven force, \textit{$\mathcal{L}_{1}$}, greatly improves the space exploration of the landmark detection as compared with \textit{$BK$}. Although the loss \textit{$\mathcal{L}_{1}$} can effectively spread the landmarks in 3D space, they are evenly distributed which cannot match the shape of the target organ (see Fig. \ref{fig:heart_kp}(b)). On this basis, the temporal coherence motion loss, \textit{$\mathcal{L}_{2}$}, further guides the model to explore the regions that may provide more effective guidance to the motion estimation. Finally, to make the detected landmarks conform to the anatomical topology, we add the reference topological loss,  \textit{$\mathcal{L}_{3}$}, which achieves the best quantitative and qualitative performance in all settings under comparison as shown in Table \ref{table:1}, Fig. \ref{fig:heart_kp}(d), and Fig. \ref{fig:heart_kp}(e). Our detected landmarks exhibit strong distribution consistency with the reference landmarks which are extracted from the manual segmentation annotations. This also reflects that the detected landmarks by our unsupervised landmark detection network can conform to the anatomical topology of the target organ.

	\begin{table}[!h]
		\centering
		\color{black}\caption{\textcolor{black}{The quantitative performance of spatiotemporal motion field estimation on the \textit{4D-C-CT} dataset with different number of landmarks.}}
		\setlength{\tabcolsep}{1.2mm}{
			\begin{tabular}{c|cccc|cccc}
				\hline
				{} &  \multicolumn{4}{c|}{\textit{$Baseline + \mathcal{L}_{1} + \mathcal{L}_{2}$}} & \multicolumn{4}{c}{$Baseline + \mathcal{L}_{1} + \mathcal{L}_{2} + \mathcal{L}_{3}$}\\
				\hline
				{}  & DSC  & HD & ASD & RVD  & DSC   & HD & ASD & RVD\\
				\hline
				12 Landmarks   &  0.722 &  11.51  & 4.29 & 0.895  & 0.784 &  8.97 & 2.66 & 0.471\\
				16 Landmarks   &  0.772 &  9.26 & 3.18 & 0.618  & 0.813 &  8.02 & 2.28 & 0.287\\
				20 Landmarks  &  0.814  & 8.25  & 2.32 & 0.397 & 0.847  & 7.13 & 2.06 & 0.263\\
				24 Landmarks   &  0.842 & 7.19  & 2.34 & 0.261 &  \textbf{0.882} & \textbf{6.16} & \textbf{1.66} & \textbf{0.192} \\
				\hline
		\end{tabular}}
		\label{table:kp}
	\end{table}

	In addition to verifying the proposed losses, we also evaluate the effects of different numbers of landmarks and report the results in Table \ref{table:kp}. We use 12, 16, 20 and 24 landmarks to represent the target organ, respectively. Note that, to effectively evaluate the impact of the numbers of landmarks on motion estimation, we leverage \textit{$BK + \mathcal{L}_{f} + \mathcal{L}_{b}$} and omit \textit{$\mathcal{L}_{o}$} to constrain the landmark detection and motion reconstruction. As expected, the more landmarks result in higher accuracy of the estimated motion fields (\textit{12-landmarks: DSC=0.722, HD=11.51; 16-landmarks: DSC=0.772, HD=9.26; 20-landmarks: DSC=0.814, HD=8.25; 24-landmarks: DSC=0.842, HD=7.19}). The 24-landmark set achieves the best results in our experiments. We have used 24 landmarks in our study and achieved the state-of-the-art results. Nevertheless, we suggest that a higher number will further improve the performance, albeit incrementally. 

	\section{Conclusion}
	\subsection{Summary}
	We present a novel unsupervised topology extraction guided motion estimation method for 4D dynamic medical images. \textcolor{black}{Our main findings are that our \textit{DSD} method: 1) effectively captures the sparse topological anatomy of the volumetric image through unsupervised landmark detection with multiple losses, especially the landmark topological loss, which aligns the detected landmarks to conform to the anatomical prior; 2) is able to better reconstruct the dense deformation field, with motion tracking guided by the detected landmarks.}
	
	
	In our framework, landmark detection can be regarded as a process of sparse encoding of dense image representation, while dense motion reconstruction is a process of decoding from sparse representation. As shown in Fig. \ref{fig:heart_kp}, we have found that even if the landmarks are not directly related to the shape of the target organ, the motion field between the paired input images can be relatively accurate (see Fig.\ref{fig:heart_kp}(b)). We attribute this to the network training where the conversation between the image coding (landmark detection) and the motion decoding (dense motion reconstruction) is established, even if the result of image coding (detected landmarks) is not morphologically meaningful. However, this compulsory learning limits the accuracy of the motion estimation. In contrast, our unsupervised landmark detection network can effectively capture the sparse landmarks to represent the obvious anatomy of target organ. This ensures that the motion field estimated from the detected landmarks is more reasonable. 
	
	
	In the comparison with the refinement scheme \textit{VoxelMorph + Demons} and directly derive the motion field by \textit{Demons} (iteratively derivation), our \textit{DSD + Demons} achieves a better performance across all metrics on both of \textit{4D-C-CT} dataset and \textit{4D-L-CT} dataset (Table \ref{table:3} and Table \ref{table:4}). This is attributed to the fact that these methods, \textit{VoxelMorph} and \textit{Demons}, estimate the deformation only by calculating the similarity of local appearance while ignoring the global structural changes. This may lead to their inability to accurately estimate the moving direction of the target organ. Thus, this is prone to incorrect estimation and makes the subsequent optimization unable to further improve the estimation (see Fig. \ref{fig:8} and Fig. \ref{fig:9}). In contrast, the initial motion field estimated from our \textit{DSD} can effectively handle the large deformation and keep the correct anatomical shape under the guidance of the whole anatomical topology change. Then, our initial estimation can be further iteratively optimized to improve the detailed estimation accuracy. In addition, our ablation study further exemplifies that our unsupervised 3D landmark detection network has high consistency and anatomical significance in temporal landmark detection even when the target object shape changes greatly (see Fig. \ref{fig:heart_kp}). 
	
	\textcolor{black}{In conclusion, we introduce a novel learning-based method to estimate the motion for the dynamic medical images. We present an unsupervised 3D landmark detection method. Particularly, our proposed topological constraint can inject prior knowledge of human anatomies into landmark detection. This strategy can make the detected landmarks conform to the anatomical prior of the organ. In addition, our proposed landmark detection network is adaptable to manual landmarks, implying that the landmark detection network can be trained leveraging the manually annotated landmarks in supervised or weakly-supervised manner.} \textcolor{black}{We note that, based on our experiment, our proposed self-supervised pre-training strategy has a positive contribution to the model’s convergence, yet no significant impact upon the accuracy of motion estimation.}

	\subsection{Limitations}
	Several limitations in our work should be discussed. First, the location of each landmark is represented by an independent image-wise heatmap. This leads to a huge requirement in GPU memory usage when the number of landmarks is increased. Next, although our method can improve the accuracy of large motion estimation, \textcolor{black}{it is not sensitive to detailed structure changes, thus requiring further refinement from conventional methods.} In the future, we suggest improving the method by reducing the GPU memory. Therefore, better estimation results can be potentially obtained by increasing the number of landmarks.

	\bibliographystyle{IEEEtran}
	\bibliography{egbib}
\end{document}